\documentclass[%
 reprint,
 amsmath,amssymb,
 aps,
]{revtex4-2}

\usepackage{graphicx}% Include figure files
\usepackage{dcolumn}% Align table columns on decimal point
\usepackage{bm}% bold math
\begin{document}

\preprint{APS/123-QED}

\title{Anisotropic field ionization in nano-clusters mediated by Brunel-electron driven plasma waves}% Force line breaks with \\

\author{Xiaohui Gao}
  \email{gaoxh@utexas.edu}
\affiliation{Department of Physics, Shaoxing University, Shaoxing, Zhejiang 312000, China}

\date{\today}% It is always \today, today,

\begin{abstract}
Ionization is one of the most fundamental processes in intense laser-matter interaction. It is extremely efficient for clusters in laser fields and often leads to surprisingly high charge states at moderate laser intensities. Here we reveal a novel ionization mechanism in laser-cluster interaction through particle-in-cell simulations. As the laser field ionizes a cluster, Brunel electrons pushed back into the clustered plasma form an attosecond bunches, impulsively exciting plasma oscillation. The resulting localized wake field further ionizes the cluster, causing a highly ionized rod-like core along the polarization axis. This anisotropic ionization is prominent using few-cycle pulses and may be washed out using longer pulses due to collisional ionization. This newly identified ionization channel can potentially provide complicated site-specific control of high ionization in nanometer-scale targets.
\end{abstract}

\maketitle

Laser-cluster interaction has been a subject of strong scientific interest for several decades~\cite{Krainov2002PR, Fennel2010RMP, Ostrikov2016RMP}. The localized high density in connection with the isolated environment enables efficient laser-matter coupling, providing not only a unique platform for studying the ultrafast non-equilibrium dynamics~\cite{Fennel2010RMP} but also various practical applications ranging from particle accelerators~\cite{Fukuda2009PRL, Matsui2019PRL} to neutron sources~\cite{Ditmire1999N} and nuclear isomers sources~\cite{Feng2022PRL}. 

A fascinating aspect of intense laser-cluster interaction is the surprisingly efficient ionization and heating. A deep understanding of the ionization dynamics is of intrinsic interest to the high-field physics community, and a convenient control of this process is of practical relevance for applications such as particle acceleration by Coulomb explosion. As the first step in intense laser-cluster interaction, ionization has been investigated extensively. It is generally divided into two categories, i.e., field ionization and collisional ionization. Efficient field ionization in clusters is often due to field enhancement, which may originate from field resonance at the critical density~\cite{Milchberg2001PRE,Koller1999PRL} or field amplification due to cluster polarization~\cite{Jungreuthmayer2004PRL}. The ionizing electric field is not limited to laser field and can be the electric field due to charge separation such as hot electron bunches~\cite{Psikal2011NIMPRA} or the free plasmon oscillations of clustered plasma driven by few-cycle pulses~\cite{Bystrov2009PRL,Gao2019OL} and the sheath electric field generated by a hot electron population in a clustering gas jet~\cite{McCormick2014PRL}. Collisional ionization by energetic electrons is another important contribution for the high ionization in clusters~\cite{Ditmire1996PRA}, but it favors longer pulses and plays a minor role for extremely short pulses~\cite{Fourment2018PRB}. The contribution of different ionization channel depends on the laser parameters and cluster properties such as the wavelength and the size~\cite{Saalmann2006JPB, Park2022PRL}, and the physical picture may not yet be exhaustive. 

While many interesting phenomena of laser-cluster interaction stem from the resonant behavior of the electrons confined within the cluster, the role of Brunel electrons is less explored. Some electrons near the surface may be pulled out of the cluster and pushed back into the plasma as the laser field decreases. These electrons known as Brunel electrons are responsible for the efficient heating in clusters~\cite{Taguchi2004PRL, Breizman2005PP}. Brunel electrons are well studied in laser interaction with solid-density bulk plasmas, uncovering a multitude of intriguing phenomena such as vacuum heating~\cite{Brunel1987PRL} and coherent wake emission~\cite{Quere2006PRL, Thaury2010JPBAMOP}.  

In this paper, we study the ionization dynamics of rare-gas clusters in ultrashort near-infrared laser pulses at moderate intensities via particle-in-cell simulations. Under the irradiation of few-cycle pulses, the charge state distribution is highly non-uniform and exhibits a pronounced rod-like hot region along the polarization axis. This enhanced ionization is due to electric field of plasma oscillation driven by Brunel-electron bunches, similar to the process in the overdense plasmas giving rise to coherent wake emission. We also find that collisional ionization is detrimental to this exotic charge distribution, and the hot spot at the center disappears with longer pulses. 

Our simulations are performed using an open-source three-dimensional particle-in-cell code \textsc{Smilei}~\cite{Derouillat2018CPC}.  The dimension of the simulation box is $6\lambda\times0.4\lambda\times0.4\lambda$, where $\lambda$ is the wavelength. The grid resolution is 320 cells per wavelength, and the time resolution is 580 steps per cycle. A cluster sits at the center of the simulation box with 125 unionized macro-particles in each cell. The laser pulse enters the simulation box along the $x$-axis from the left. The pulse is linearly polarized in the $y$ direction and is assumed to be a plane wave in space and a Gaussian profile in time. The field employs silver-muller boundary conditions for the longitudinal direction and periodic boundary conditions for the transverse directions. Field ionization, electron-electron collision, electron-ion collision, and collisional ionization are included in the simulation using standard modules of the code. Recombination is not accounted for, which should not alter the dynamics significantly in the few-cycle regime. The polarizability of the neutral atom is neglected, which is justified when the size is much smaller than the wavelength or the intensity is substantially higher than the ionization threshold~\cite{Gao2023PP}. In all of our simulations, we consider a spherical argon cluster interacting with a 800-nm pulse. The cluster radius is 40-nm, which is large enough to capture the dynamics due to plasma wave as the size is several times the plasma oscillation wavelength. Such clusters are routinely produced in experiments by pulsed gas jets backed with high-pressure gases. The atomic density of argon cluster is $2.6\times10^{22}$\,cm$^{-3}$~\cite{Gao2013JAP}, which corresponds to $15n_c$ with $n_c$ the critical density at 800-nm. 
 
\begin{figure}[htbp]
\centering
\includegraphics[width=0.45\textwidth]{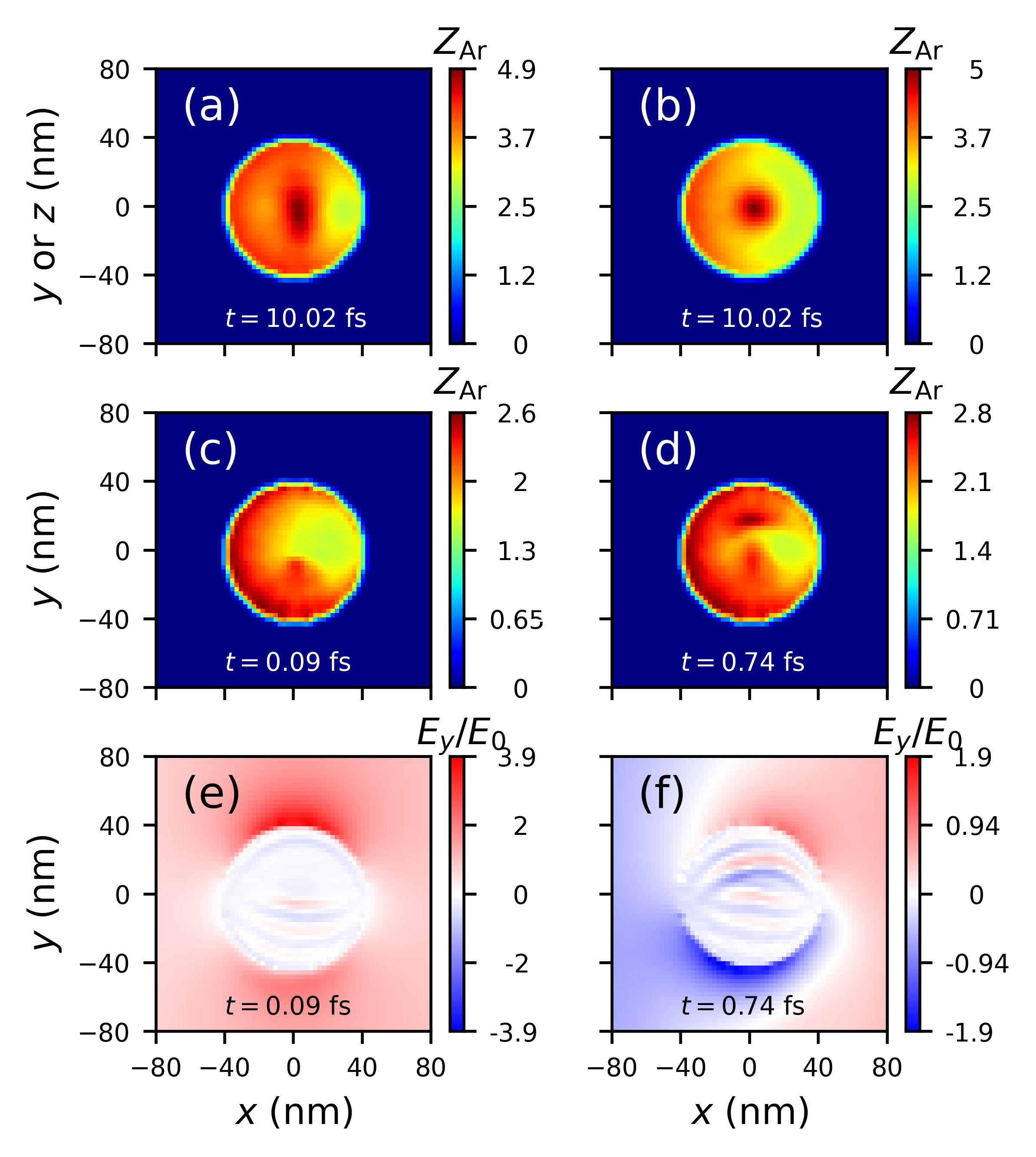}
\caption{Snapshots of an argon cluster subject to a two-cycle (FWHM in field) 800-nm pulse. (a) and (b) respectively show the charge state distribution in the $y$-$x$ plane and in the $z$-$x$ plane at $t=10.02$\,fs. (c) and (d) respectively show the charge state distribution in the $y$-$x$ plane at $t=0.09$ fs and $t=0.74$ fs. (e) and (f) respectively show the transverse field $E_y$ in the $y$-$x$ plane at $t=0.09$ fs and $t=0.74$ fs. The electric field is normalized by the driving laser field $E_0$, and $t=0$ marks the time at which the peak of the pulse reaches $x=0$.}
\label{fig1}
\end{figure}%
Figure~\ref{fig1} shows snapshots of the interaction of a 40-nm argon cluster with an intense ultrashort pulse at different observation times. The peak intensity is $5.0\times10^{15}\,$W/cm$^2$, the full width at half maximum (FWHM) of the field is two-cycle with zero carrier-envelope phase. Figure~\ref{fig1}(a) and \ref{fig1}(b) present the charge state distribution in the $x$-$y$ plane and $x$-$z$ plane, respectively, at a time of observation $t=10.02\,$fs, at which time the pulse has already passed the cluster.
An enhanced ionization is observed at the cluster center with a rod-like hot region orientated along the polarization axis. This is in sharp contrast with a general perception that ionization occurs preferentially at the periphery of the cluster due to the skin effect. The skin depth for a plasma in a laser at a non-resonant wavelength is $\delta\approx c/\omega_p$, where $c$ is the speed of light in vacuum, and $\omega_p$ is the plasma frequency. This corresponds to $\delta\approx19\,$ nm for an average charge state of three. The formation dynamics of this unexpected ionization pattern is illustrated in Figs.~\ref{fig1}(c)-(f). The initial ionization is due to the propagation of the laser field and is higher in the front surface.  An ionization wave, however, propagates from the bottom pole toward the center, as shown in the snapshot of charge state distribution at $t=0.09\,$fs in [Fig.~\ref{fig1}(c)]. The snapshot of charge state distribution at $t=0.74\,$fs in Fig.~\ref{fig1}(d) shows another ionization wave propagating from the top pole toward the center. The ionization is consistent with the corresponding field distribution $E_y$ shown in Fig.~\ref{fig1}(e) and Fig.~\ref{fig1}(f). 

\begin{figure}[htbp]
\centering
\includegraphics[width=0.45\textwidth]{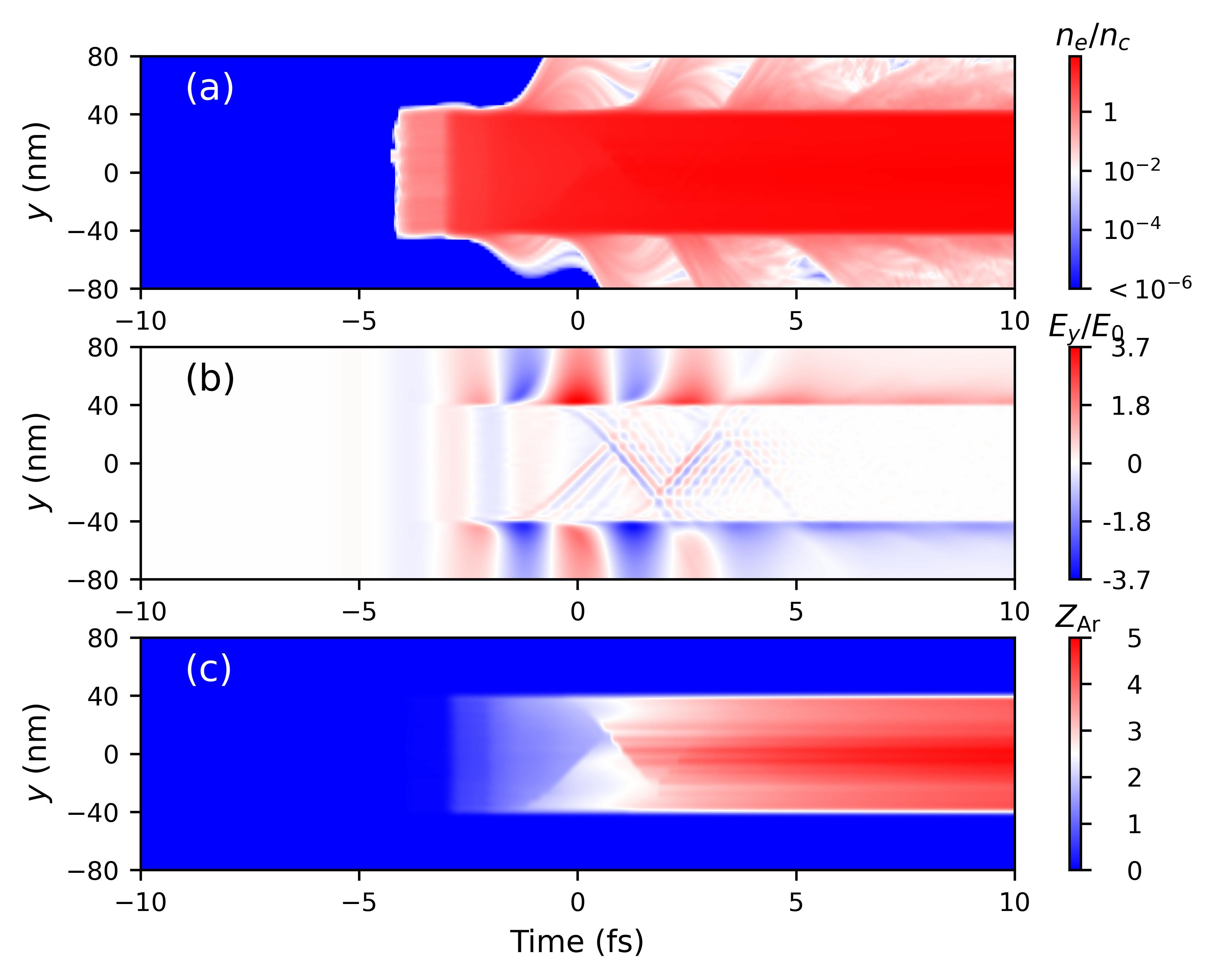}
\caption{ Time evolution of three quantities at the polarization axis ($x,z)=(0,0$). (a) Electron density $n_e$ normalized by the critical density $n_c$. (b) Transverse field $E_y$ normalized by the driving field $E_0$. (c) Charge state $Z$ of the argon atoms}
\label{fig2}
\end{figure}%
To clarify the underlying ionization mechanism, streaks of the electron density $n_e$, transverse field $E_y$, and charge state $Z_\mathrm{Ar}$ along the polarization axis are presented against the observation time. 
Figure~\ref{fig2}(a) shows that a small fraction of electrons is pulled into vacuum by the laser field and is then sent back into the cluster. This occurs alternatingly at the top pole and at the bottom pole every half-cycle. As Brunel electrons with a longer trajectory gain more energy than those with a shorter trajectory, trajectory crossing occurs inside the cluster and forms an attosecond electron bunch with a appreciable electron density peak, exciting collective plasma oscillations of frequency $\omega_p$ in its wake. The analysis of plasma oscillation excited by Brunel electrons is similar to the process in plasma mirrors~\cite{Quere2006PRL, Thaury2010JPBAMOP} except for two distinct features. First, the plasma wave in clusters has a curvature due to the spherical geometry, leading to an increased oscillation amplitude as  the wave converges. Secondly, the waves are excited twice in every cycle in clusters as Brunel electrons are produced both at the top and bottom part of the clusters. The evolution of the plasma wave can be seen from Fig.~\ref{fig2}(b), where the plasma oscillations are damped quickly, and oscillations excited from different bunches have interference. The ionization front shown in Fig.~\ref{fig2}(c) occurs at the same time of the plasma wave, confirming that the Brunel-electron driven plasma wave is responsible for the enhanced ionization at the core of clusters.

\begin{figure}[htbp]
\centering
\includegraphics[width=0.35\textwidth]{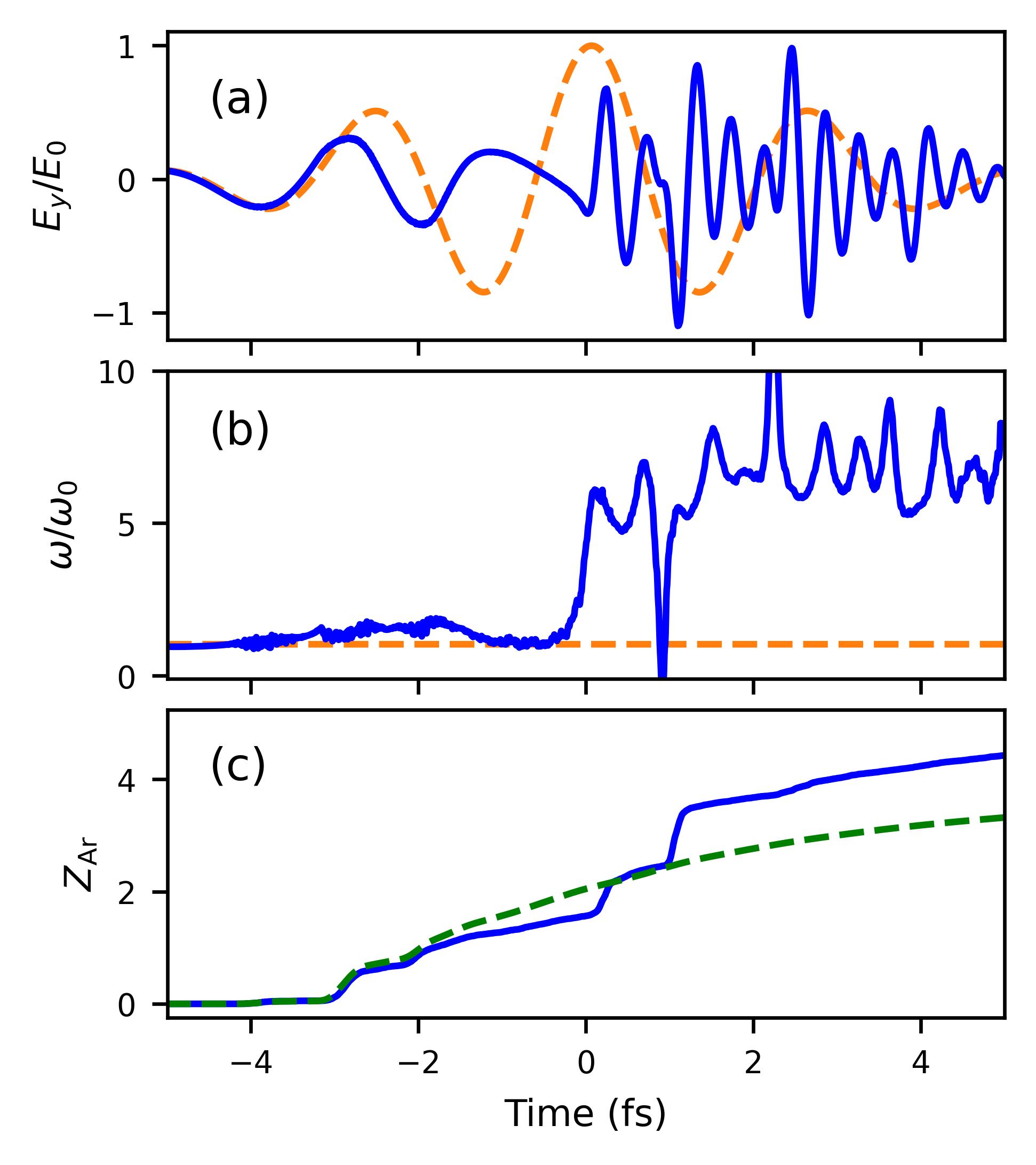}
\caption{(a) Normalized electric field $E_y/E_0$ at the center of the cluster as a function of time. (b) The corresponding instantaneous angular frequency $\omega$ normalized by the laser angular frequency $\omega_0$. The dashed orange curves show the case when a cluster is not present. (c) Charge state $Z$ at the center (solid blue curve) and charge state averaged over the entire cluster (dashed green curve) as a function of time.}
\label{fig3}
\end{figure}%
Figures~\ref{fig3}(a) and ~\ref{fig3}(b) respectively show the temporal electric field $E_y(t)$ and the corresponding instantaneous angular frequency $\omega(t)$ calculated as the temporal derivative of the oscillation phase of the electric field at the cluster center. Figure~\ref{fig3}(c) shows the time evolution of the charge state at the cluster center (solid blue curve) and the charge state averaged over the entire cluster (dashed orange curve). The field is initially shielded due to ionization at the surface. At approximately $t=0$, the field starts to oscillate at roughly six times that of the laser frequency, and the charge state at the center rises steeply. This corresponds to the time that the first electron bunch travels to the cluster center. 
Note that the electron density is approximately $30n_c$. Thus the field oscillates roughly at $\omega_p$ rather than the surface plasmon frequency $\omega_p/\sqrt{3}$. The charge state exhibits a second steep rise approximately at $t=0.9\,$fs, at which time the instantaneous frequency exhibits a sharp dip. This is caused by the interference of plasma oscillations excited by the electron bunch entered from the bottom pole and that from the top pole. After $t=0.9$\,fs, the charge state at the cluster center is substantially higher than the average charge state, indicating a formation of a highly ionized core.

\begin{figure}[htbp]
\centering
\includegraphics[width=0.4\textwidth]{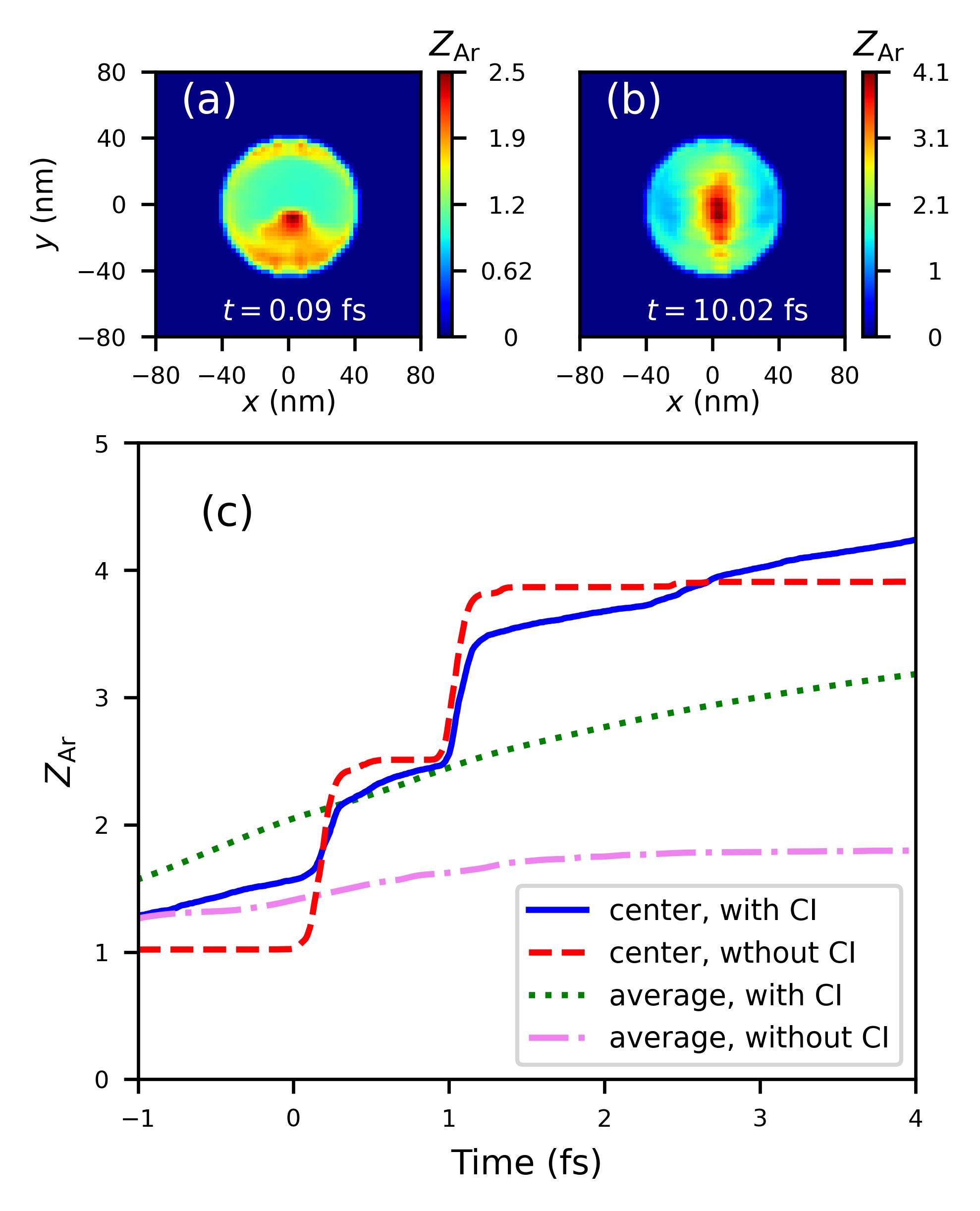}
\caption{Charge state distribution in the $y$-$x$ plane at two observation times, (a) $t=0.09$\,fs, (b) $t=10.02$\,fs. (c) Time evolution of charge state at the center of the cluster and the charge state averaged over the entire cluster with and without collisional ionization (CI).}
\label{fig4}
\end{figure}% 
Collisional ionization is the other ionization channel when a fast electron bunch travels in a partially ionized plasma.  To evaluate its role, collisional ionization is switched off and the resulting charge distribution at two observation times $t=0.09$\,fs and $t=10.02$\,fs are shown in the Fig.~\ref{fig4}(a) and Fig.~\ref{fig4}(b), respectively. With field ionization alone, an even more pronounced anisotropic ionization is observed. The temporal variations of $Z_\mathrm{Ar}$ at the center and $Z_\mathrm{Ar}$ averaged over the entire cluster for cases with and without collisional ionization are presented in Fig.~\ref{fig4}(c). In spite of the few-cycle pulse duration, collisional ionization significantly increases the average charge and causes further ionization when the pulse is gone ($t>2$\,fs). Interestingly, the collisional ionization brings the charge state at the center slightly down between $t=0$ and $t=2\,$ fs. The reason could be that the collisional ionization dissipates the energy of driving electron bunches, leaving an inefficient excitation of plasma oscillation.

\begin{figure}[htbp]
\centering
\includegraphics[width=0.4\textwidth]{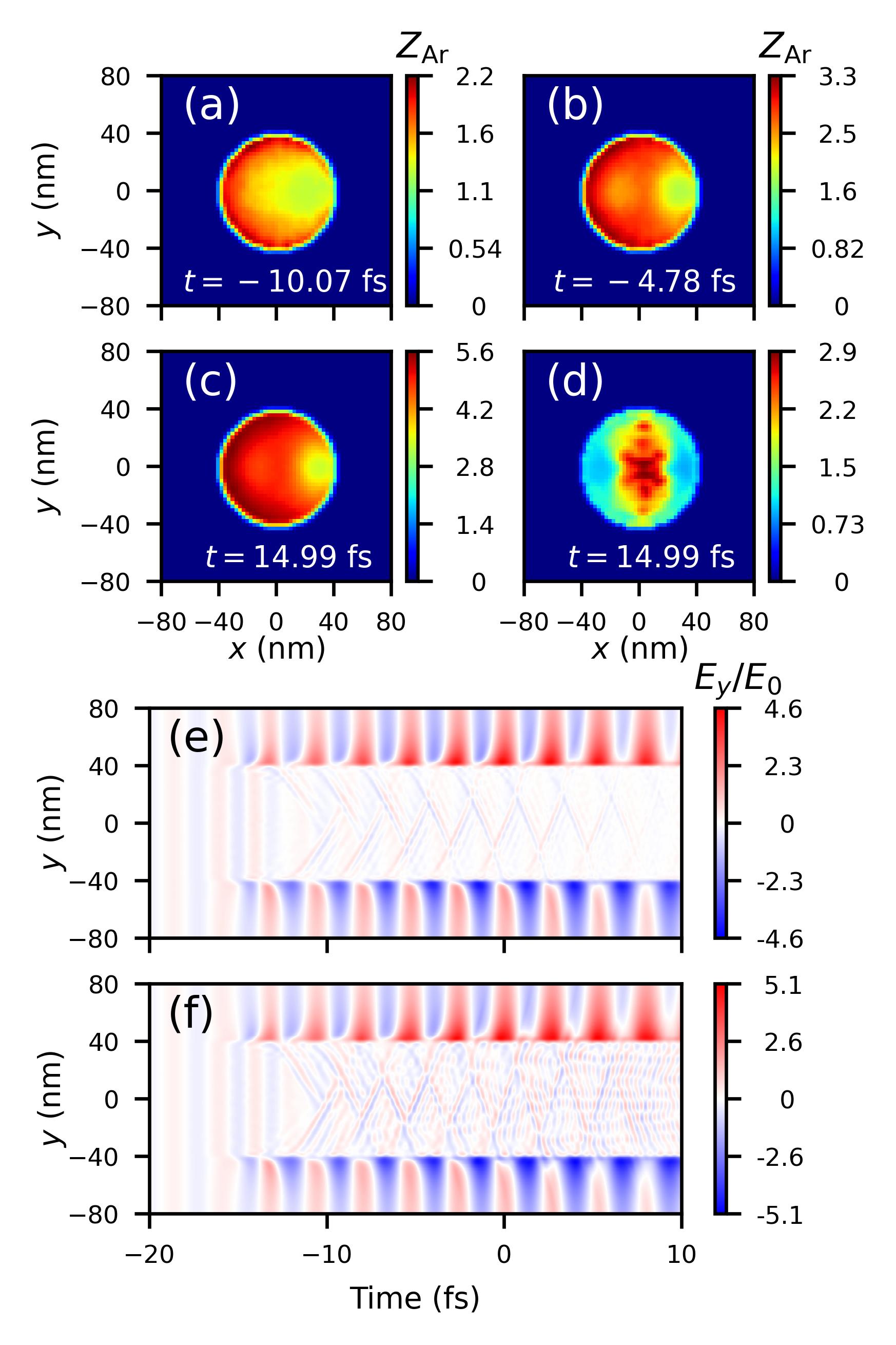}
\caption{Results of a 40-nm argon cluster interacting with a 10-cycle (FWHM in field) pulse of $2.0\times10^{15}\,$W/cm$^2$. Charge state distribution in the $y$-$x$ plane at three observation times: (a) $t=-10.07$\,fs, (b) $t=-4.78$\,fs, (c) $t=14.99$\,fs. (d) Charge state distribution in the $y$-$x$ plane at $t=14.99$\,fs when collisional ionization is switched off. (e) Time evolution of transverse field $E_y$ at polarization axis ($x,z)=(0,0$). (f) The case when collisional ionization is switched off.}
\label{fig5}
\end{figure}% 
Because of the negative role of the collisional ionization in the plasma oscillation, it is expected that enhanced core ionization may disappear at longer pulses. This is confirmed in the simulation for a 10-cycle pulse at $2.0\times10^{15}\,$W/cm$^2$ peak intensity. Figure~\ref{fig5}(a) shows that ionization starts at the front surface of the cluster, and Fig.~\ref{fig5}(b) shows a weak column-like structure along the polarization axis is developed. However, at $t=14.99\,$fs, the ionization is almost homogeneous except for the surface [Fig.~\ref{fig5}(c)]. A completely different pattern is observed when the collisional ionization is turned off [Fig.~\ref{fig5}(d)]. Figures~\ref{fig5}(e) and ~\ref{fig5}(f) show the time evolution of the field $E_y$ along polarization axis for the case with and without collisional ionization, respectively. The internal field is weaker with the collisional ionization, leaving the cluster core less ionized.

\begin{figure}[htbp]
\centering
\includegraphics[width=0.4\textwidth]{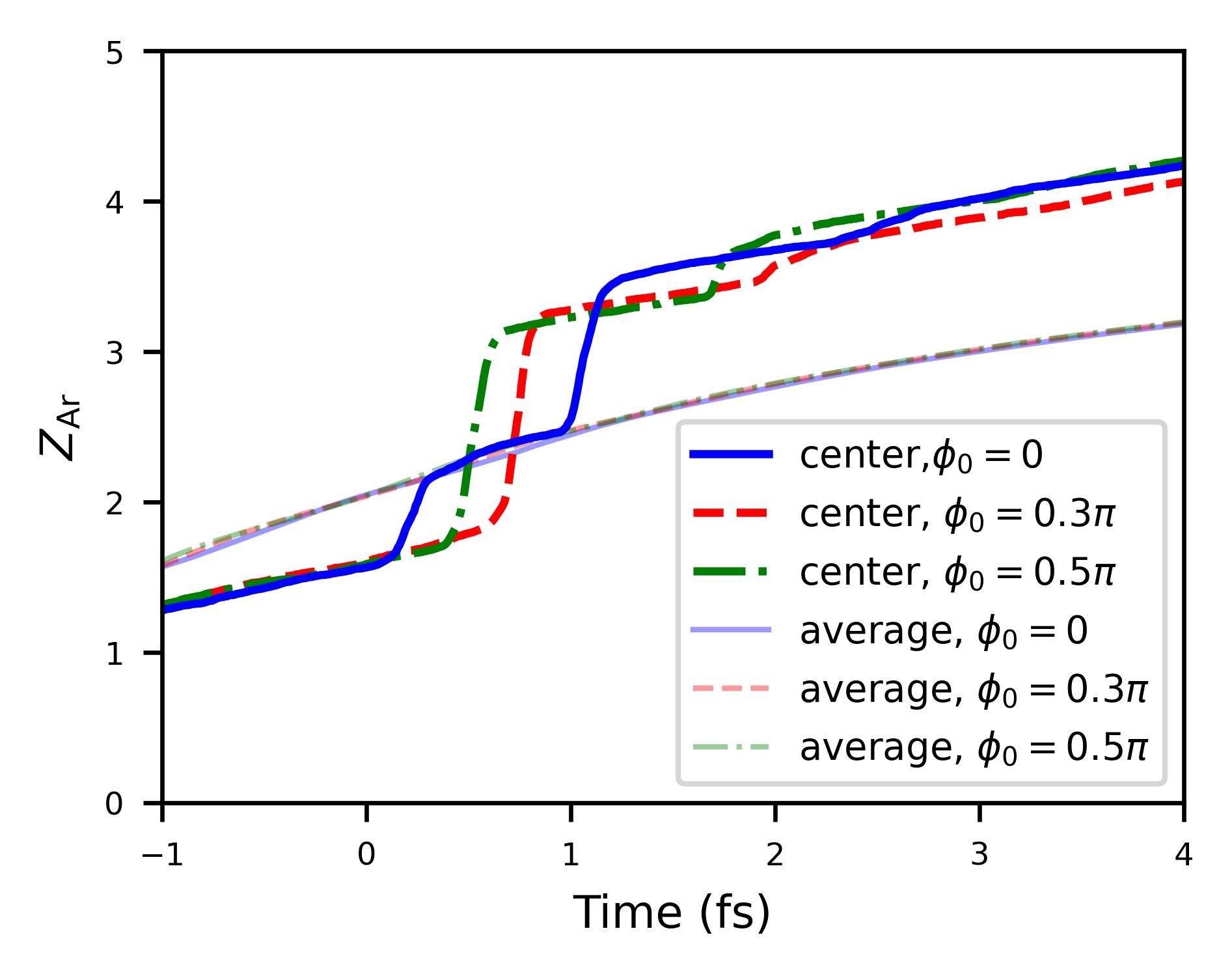}
\caption{Time evolution of charge state at the center of the cluster and the charge state averaged over the entire cluster for three different carrier-envelope phases: 0, 0.3$\pi$, and 0.5$\pi$.}
\label{fig6}
\end{figure}%  

As this ionization mechanism stems from the subcycle dynamics of the Brunel electrons and works best with few-cycle pulses, the role of the carrier-envelope phase deserves attention. Figure~\ref{fig6} shows the variations of charge state at the cluster center and the volume-averaged charge state for three different carrier-envelope phase. The solid blue lines represent the results from the same simulation as in Figs.~\ref{fig1}, where $\phi_0$=0 is used. The dashed red line and dash-dotted green line represent the cases with $\phi_0=0.3\pi$ and $\phi_0=0.5\pi$, respectively. The subcyle evolution of the charge state at the cluster center is quite different, but the final charge state at the center, which is a time-integrated effect, shows little variation with the carrier-envelope phase. The volume averaged charge state is almost identical for the three cases at all the time.  

\begin{figure}[htbp]
\centering
\includegraphics[]{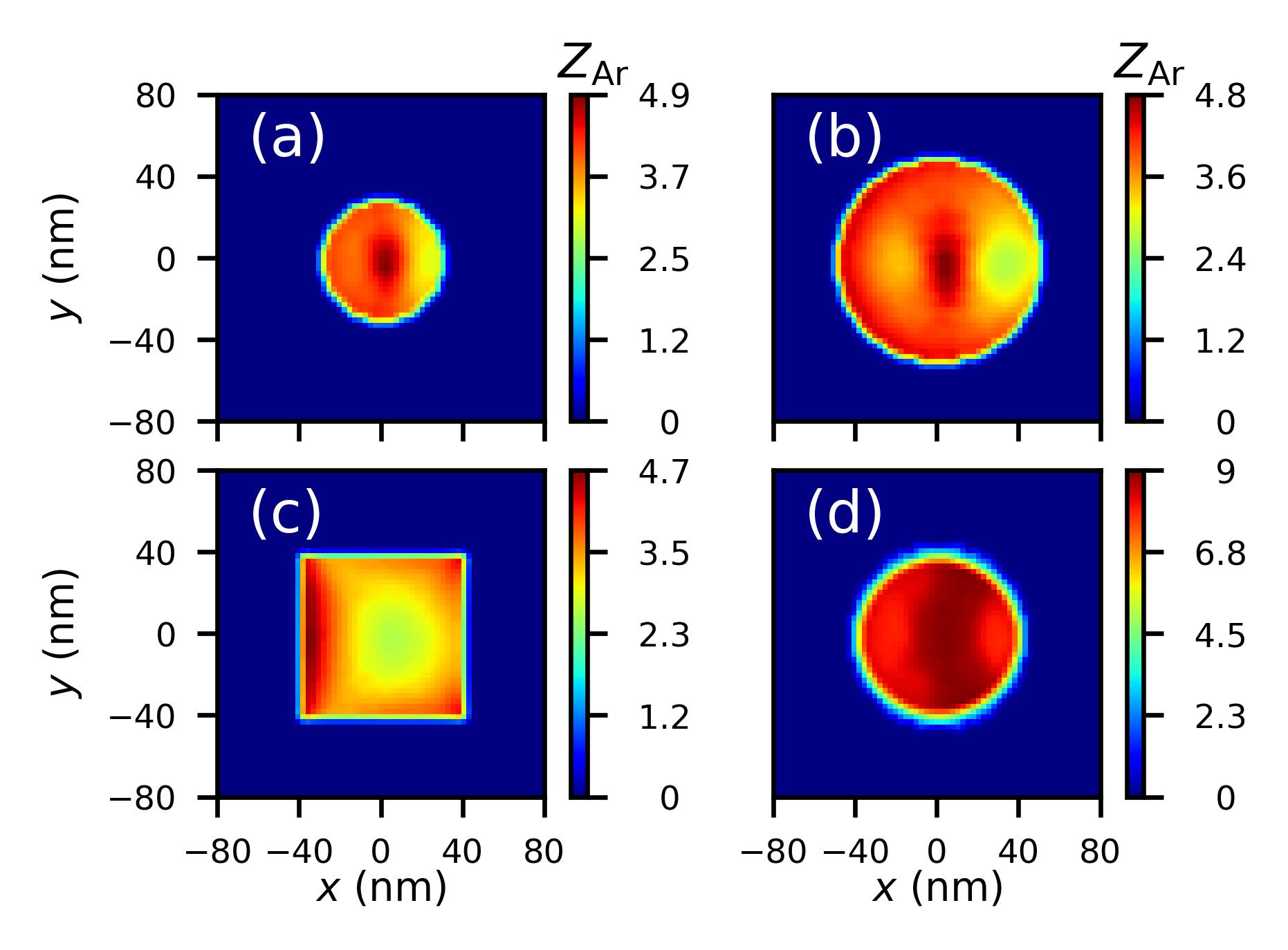}
\caption{Charge state distribution in the $y$-$x$ plane at $t=10.02$\,fs for (a) a spherical cluster of 30-nm radius, (b) a spherical cluster of 50-nm radius, (c) a phantom cubic cluster of 80-nm, (d) a 40-nm cluster with an intensity of $5\times10^{17}\,$W/cm$^2$. Other unspecified parameters are the same as in Fig.~\ref{fig1}.}
\label{fig7}
\end{figure}% 

In additional to the pulse duration and carrier-envelope phase, we also investigate other configures. Figure~\ref{fig7}(a) and \ref{fig7}(b) show the snapshot of the charge state distribution at $t=10.02$\,fs for 30-nm and 50-nm sized clusters. The pattern is similar, although a larger size produces a slightly lower maximum charge state at the center. This mechanism is necessarily associated with the curvature of the cluster surface. For a phantom cubic cluster as shown in Fig.~\ref{fig7}(c), no core ionization is observed. Without the focusing of the wave due to the spherical geometry, the field is not strong enough to create appreciable ionization. Figure~\ref{fig7}(d) shows the case with a much higher intensity of $5\times10^{17}$\,W/cm$^2$. Due to big gap of ionization potential for Ar$^{8+}$, the ionization is saturated to some degree, leading to a smaller anisotropy. 

Our mechanism shares the two steps of the semiclassical three-step model of high harmonic generation~\cite{Corkum1993PRL}, which are the ionization and propagation. However, the last step of three-step model is a collisional process while our scenario relies on the field from the Brunel-electron-driven plasma oscillation. This novel ionization mechanism has not received attention before probably due to the following reasons. First, quite a few studies consider micrometer-sized hydrogen clusters or helium droplets~\cite{Liseykina2013PRL, Zastrau2015JPBAMOP}. As the core of these large clusters remains unionized, the plasma-wave driven ionization is absent. Second, this effect manifests itself in the few-cycle regime, which is less explored because the majority of terawatt laser systems generate pulse of several tens of femtosecond duration. Third, study of laser-cluster interaction is heavily motivated by the application of energetic particle generation, where an ultra-intense laser pulse with an intensity exceeding $10^{17}$ W/cm$^2$ is often used. At such an intensity, the ionization is complete at the leading edge of the pulse and the ionization dynamics is not of particular interest.
Our scenario bears some similarities with the previously studied attosecond plasma wave dynamics in nanoplasmas~\cite{Varin2012PRL}. However, resonant conditions are required in their scheme, which is met by using a pump-probe scheme. The resonant excitation causes a field enhancement of several orders of magnitude and alters the dynamics of collisional ionization, allowing a strongly nonuniform ion charge distribution at longer pulse duration. Ionization hotspot at the cluster core was also demonstrated for a helium nanodroplet doped with a tiny xenon cluster ~\cite{Mikaberidze2009PRL}. Recently, it is found that clusters pumped with longer wavelength produce a column-like inhomogeneous charge distribution. However, cluster poles remain the preferred sites for ionization~\cite{Park2022PRL}.

%While ionization dynamics in clusters or droplets has been investigated for a time, most studies consider micrometer-sized hydrogen clusters or helium droplets. As the core of these large clusters remains unionized, the plasma-wave driven ionization is absent. 

In summary, we have identified a novel ionization mechanism for nano-clusters in intense ultrashort laser pulses using particle-in-cell simulations. In a partially ionized cluster, the returning Brunel electrons form an attosecond bunch and drive plasma oscillation impulsively. The resulting electric field causes higher ionization. This scheme is favored at shorter pulse duration where collisional ionization plays a minor role. With the increasing availability of intense few-cycle laser sources and the advancement of the single-shot coherent X-ray diffractive imaging techniques~\cite{Gorkhover2016NP, Bacellar2022PRL}, it is feasible to test this anisotropic ionization experimentally. Our study will stimulate further experiment investigation, and may offer a convenient way to control the ionization and heating in nanometer-sized targets by using few-cycle pulses with an engineered curvature of target surface.

\begin{acknowledgments}
This work was supported by Natural Science Foundation of Zhejiang Province (LY19A040005).
\end{acknowledgments}

%\nocite{*}


\begin{thebibliography}{33}%
\makeatletter
\providecommand \@ifxundefined [1]{%
 \@ifx{#1\undefined}
}%
\providecommand \@ifnum [1]{%
 \ifnum #1\expandafter \@firstoftwo
 \else \expandafter \@secondoftwo
 \fi
}%
\providecommand \@ifx [1]{%
 \ifx #1\expandafter \@firstoftwo
 \else \expandafter \@secondoftwo
 \fi
}%
\providecommand \natexlab [1]{#1}%
\providecommand \enquote  [1]{``#1''}%
\providecommand \bibnamefont  [1]{#1}%
\providecommand \bibfnamefont [1]{#1}%
\providecommand \citenamefont [1]{#1}%
\providecommand \href@noop [0]{\@secondoftwo}%
\providecommand \href [0]{\begingroup \@sanitize@url \@href}%
\providecommand \@href[1]{\@@startlink{#1}\@@href}%
\providecommand \@@href[1]{\endgroup#1\@@endlink}%
\providecommand \@sanitize@url [0]{\catcode `\\12\catcode `\$12\catcode
  `\&12\catcode `\#12\catcode `\^12\catcode `\_12\catcode `\%12\relax}%
\providecommand \@@startlink[1]{}%
\providecommand \@@endlink[0]{}%
\providecommand \url  [0]{\begingroup\@sanitize@url \@url }%
\providecommand \@url [1]{\endgroup\@href {#1}{\urlprefix }}%
\providecommand \urlprefix  [0]{URL }%
\providecommand \Eprint [0]{\href }%
\providecommand \doibase [0]{https://doi.org/}%
\providecommand \selectlanguage [0]{\@gobble}%
\providecommand \bibinfo  [0]{\@secondoftwo}%
\providecommand \bibfield  [0]{\@secondoftwo}%
\providecommand \translation [1]{[#1]}%
\providecommand \BibitemOpen [0]{}%
\providecommand \bibitemStop [0]{}%
\providecommand \bibitemNoStop [0]{.\EOS\space}%
\providecommand \EOS [0]{\spacefactor3000\relax}%
\providecommand \BibitemShut  [1]{\csname bibitem#1\endcsname}%
\let\auto@bib@innerbib\@empty
%</preamble>
\bibitem [{\citenamefont {Krainov}\ and\ \citenamefont
  {Smirnov}(2002)}]{Krainov2002PR}%
  \BibitemOpen
  \bibfield  {author} {\bibinfo {author} {\bibfnamefont {V.~P.}\ \bibnamefont
  {Krainov}}\ and\ \bibinfo {author} {\bibfnamefont {M.~B.}\ \bibnamefont
  {Smirnov}},\ }\bibfield  {title} {\bibinfo {title} {Cluster beams in the
  super-intense femtosecond laser pulse},\ }\href
  {http://www.sciencedirect.com/science/article/pii/S0370157302002727}
  {\bibfield  {journal} {\bibinfo  {journal} {Phys. Rep.}\ }\textbf {\bibinfo
  {volume} {370}},\ \bibinfo {pages} {237} (\bibinfo {year}
  {2002})}\BibitemShut {NoStop}%
\bibitem [{\citenamefont {Fennel}\ \emph {et~al.}(2010)\citenamefont {Fennel},
  \citenamefont {Meiwes-Broer}, \citenamefont {Tiggesbaumker}, \citenamefont
  {Reinhard}, \citenamefont {Dinh},\ and\ \citenamefont
  {Suraud}}]{Fennel2010RMP}%
  \BibitemOpen
  \bibfield  {author} {\bibinfo {author} {\bibfnamefont {T.}~\bibnamefont
  {Fennel}}, \bibinfo {author} {\bibfnamefont {K.~H.}\ \bibnamefont
  {Meiwes-Broer}}, \bibinfo {author} {\bibfnamefont {J.}~\bibnamefont
  {Tiggesbaumker}}, \bibinfo {author} {\bibfnamefont {P.~G.}\ \bibnamefont
  {Reinhard}}, \bibinfo {author} {\bibfnamefont {P.~M.}\ \bibnamefont {Dinh}},\
  and\ \bibinfo {author} {\bibfnamefont {E.}~\bibnamefont {Suraud}},\
  }\bibfield  {title} {\bibinfo {title} {Laser-driven nonlinear cluster
  dynamics},\ }\href {https://doi.org/10.1103/RevModPhys.82.1793} {\bibfield
  {journal} {\bibinfo  {journal} {Rev. Mod. Phys.}\ }\textbf {\bibinfo {volume}
  {82}},\ \bibinfo {pages} {1793} (\bibinfo {year} {2010})}\BibitemShut
  {NoStop}%
\bibitem [{\citenamefont {Ostrikov}\ \emph {et~al.}(2016)\citenamefont
  {Ostrikov}, \citenamefont {Beg},\ and\ \citenamefont {Ng}}]{Ostrikov2016RMP}%
  \BibitemOpen
  \bibfield  {author} {\bibinfo {author} {\bibfnamefont {K.~K.}\ \bibnamefont
  {Ostrikov}}, \bibinfo {author} {\bibfnamefont {F.}~\bibnamefont {Beg}},\ and\
  \bibinfo {author} {\bibfnamefont {A.}~\bibnamefont {Ng}},\ }\bibfield
  {title} {\bibinfo {title} {Nanoplasmas generated by intense radiation},\
  }\href {https://link.aps.org/doi/10.1103/RevModPhys.88.011001} {\bibfield
  {journal} {\bibinfo  {journal} {Rev. Mod. Phys.}\ }\textbf {\bibinfo {volume}
  {88}},\ \bibinfo {pages} {011001} (\bibinfo {year} {2016})}\BibitemShut
  {NoStop}%
\bibitem [{\citenamefont {Fukuda}\ \emph {et~al.}(2009)\citenamefont {Fukuda},
  \citenamefont {Faenov}, \citenamefont {Tampo}, \citenamefont {Pikuz},
  \citenamefont {Nakamura}, \citenamefont {Kando}, \citenamefont {Hayashi},
  \citenamefont {Yogo}, \citenamefont {Sakaki}, \citenamefont {Kameshima},
  \citenamefont {Pirozhkov}, \citenamefont {Ogura}, \citenamefont {Mori},
  \citenamefont {Esirkepov}, \citenamefont {Koga}, \citenamefont {Boldarev},
  \citenamefont {Gasilov}, \citenamefont {Magunov}, \citenamefont {Yamauchi},
  \citenamefont {Kodama}, \citenamefont {Bolton}, \citenamefont {Kato},
  \citenamefont {Tajima}, \citenamefont {Daido},\ and\ \citenamefont
  {Bulanov}}]{Fukuda2009PRL}%
  \BibitemOpen
  \bibfield  {author} {\bibinfo {author} {\bibfnamefont {Y.}~\bibnamefont
  {Fukuda}}, \bibinfo {author} {\bibfnamefont {A.~Y.}\ \bibnamefont {Faenov}},
  \bibinfo {author} {\bibfnamefont {M.}~\bibnamefont {Tampo}}, \bibinfo
  {author} {\bibfnamefont {T.~A.}\ \bibnamefont {Pikuz}}, \bibinfo {author}
  {\bibfnamefont {T.}~\bibnamefont {Nakamura}}, \bibinfo {author}
  {\bibfnamefont {M.}~\bibnamefont {Kando}}, \bibinfo {author} {\bibfnamefont
  {Y.}~\bibnamefont {Hayashi}}, \bibinfo {author} {\bibfnamefont
  {A.}~\bibnamefont {Yogo}}, \bibinfo {author} {\bibfnamefont {H.}~\bibnamefont
  {Sakaki}}, \bibinfo {author} {\bibfnamefont {T.}~\bibnamefont {Kameshima}},
  \bibinfo {author} {\bibfnamefont {A.~S.}\ \bibnamefont {Pirozhkov}}, \bibinfo
  {author} {\bibfnamefont {K.}~\bibnamefont {Ogura}}, \bibinfo {author}
  {\bibfnamefont {M.}~\bibnamefont {Mori}}, \bibinfo {author} {\bibfnamefont
  {T.~Z.}\ \bibnamefont {Esirkepov}}, \bibinfo {author} {\bibfnamefont
  {J.}~\bibnamefont {Koga}}, \bibinfo {author} {\bibfnamefont {A.~S.}\
  \bibnamefont {Boldarev}}, \bibinfo {author} {\bibfnamefont {V.~A.}\
  \bibnamefont {Gasilov}}, \bibinfo {author} {\bibfnamefont {A.~I.}\
  \bibnamefont {Magunov}}, \bibinfo {author} {\bibfnamefont {T.}~\bibnamefont
  {Yamauchi}}, \bibinfo {author} {\bibfnamefont {R.}~\bibnamefont {Kodama}},
  \bibinfo {author} {\bibfnamefont {P.~R.}\ \bibnamefont {Bolton}}, \bibinfo
  {author} {\bibfnamefont {Y.}~\bibnamefont {Kato}}, \bibinfo {author}
  {\bibfnamefont {T.}~\bibnamefont {Tajima}}, \bibinfo {author} {\bibfnamefont
  {H.}~\bibnamefont {Daido}},\ and\ \bibinfo {author} {\bibfnamefont {S.~V.}\
  \bibnamefont {Bulanov}},\ }\bibfield  {title} {\bibinfo {title} {Energy
  increase in multi-mev ion acceleration in the interaction of a short pulse
  laser with a cluster-gas target},\ }\href
  {https://doi.org/10.1103/PhysRevLett.103.165002} {\bibfield  {journal}
  {\bibinfo  {journal} {Phys. Rev. Lett.}\ }\textbf {\bibinfo {volume} {103}},\
  \bibinfo {pages} {165002} (\bibinfo {year} {2009})}\BibitemShut {NoStop}%
\bibitem [{\citenamefont {Matsui}\ \emph {et~al.}(2019)\citenamefont {Matsui},
  \citenamefont {Fukuda},\ and\ \citenamefont {Kishimoto}}]{Matsui2019PRL}%
  \BibitemOpen
  \bibfield  {author} {\bibinfo {author} {\bibfnamefont {R.}~\bibnamefont
  {Matsui}}, \bibinfo {author} {\bibfnamefont {Y.}~\bibnamefont {Fukuda}},\
  and\ \bibinfo {author} {\bibfnamefont {Y.}~\bibnamefont {Kishimoto}},\
  }\bibfield  {title} {\bibinfo {title} {Quasimonoenergetic proton bunch
  acceleration driven by hemispherically converging collisionless shock in a
  hydrogen cluster coupled with relativistically induced transparency},\ }\href
  {https://link.aps.org/doi/10.1103/PhysRevLett.122.014804} {\bibfield
  {journal} {\bibinfo  {journal} {Phys. Rev. Lett.}\ }\textbf {\bibinfo
  {volume} {122}},\ \bibinfo {pages} {014804} (\bibinfo {year}
  {2019})}\BibitemShut {NoStop}%
\bibitem [{\citenamefont {Ditmire}\ \emph {et~al.}(1999)\citenamefont
  {Ditmire}, \citenamefont {Zweiback}, \citenamefont {Yanovsky}, \citenamefont
  {Cowan}, \citenamefont {Hays},\ and\ \citenamefont {Wharton}}]{Ditmire1999N}%
  \BibitemOpen
  \bibfield  {author} {\bibinfo {author} {\bibfnamefont {T.}~\bibnamefont
  {Ditmire}}, \bibinfo {author} {\bibfnamefont {J.}~\bibnamefont {Zweiback}},
  \bibinfo {author} {\bibfnamefont {V.~P.}\ \bibnamefont {Yanovsky}}, \bibinfo
  {author} {\bibfnamefont {T.~E.}\ \bibnamefont {Cowan}}, \bibinfo {author}
  {\bibfnamefont {G.}~\bibnamefont {Hays}},\ and\ \bibinfo {author}
  {\bibfnamefont {K.~B.}\ \bibnamefont {Wharton}},\ }\bibfield  {title}
  {\bibinfo {title} {Nuclear fusion from explosions of femtosecond laser-heated
  deuterium clusters},\ }\href {http://dx.doi.org/10.1038/19037} {\bibfield
  {journal} {\bibinfo  {journal} {Nature}\ }\textbf {\bibinfo {volume} {398}},\
  \bibinfo {pages} {489} (\bibinfo {year} {1999})}\BibitemShut {NoStop}%
\bibitem [{\citenamefont {Feng}\ \emph {et~al.}(2022)\citenamefont {Feng},
  \citenamefont {Wang}, \citenamefont {Fu}, \citenamefont {Chen}, \citenamefont
  {Tan}, \citenamefont {Li}, \citenamefont {Wang}, \citenamefont {Li},
  \citenamefont {Zhang}, \citenamefont {Ma},\ and\ \citenamefont
  {Zhang}}]{Feng2022PRL}%
  \BibitemOpen
  \bibfield  {author} {\bibinfo {author} {\bibfnamefont {J.}~\bibnamefont
  {Feng}}, \bibinfo {author} {\bibfnamefont {W.}~\bibnamefont {Wang}}, \bibinfo
  {author} {\bibfnamefont {C.}~\bibnamefont {Fu}}, \bibinfo {author}
  {\bibfnamefont {L.}~\bibnamefont {Chen}}, \bibinfo {author} {\bibfnamefont
  {J.}~\bibnamefont {Tan}}, \bibinfo {author} {\bibfnamefont {Y.}~\bibnamefont
  {Li}}, \bibinfo {author} {\bibfnamefont {J.}~\bibnamefont {Wang}}, \bibinfo
  {author} {\bibfnamefont {Y.}~\bibnamefont {Li}}, \bibinfo {author}
  {\bibfnamefont {G.}~\bibnamefont {Zhang}}, \bibinfo {author} {\bibfnamefont
  {Y.}~\bibnamefont {Ma}},\ and\ \bibinfo {author} {\bibfnamefont
  {J.}~\bibnamefont {Zhang}},\ }\bibfield  {title} {\bibinfo {title}
  {Femtosecond pumping of nuclear isomeric states by the coulomb collision of
  ions with quivering electrons},\ }\href
  {https://doi.org/10.1103/PhysRevLett.128.052501} {\bibfield  {journal}
  {\bibinfo  {journal} {Phys. Rev. Lett.}\ }\textbf {\bibinfo {volume} {128}},\
  \bibinfo {pages} {052501} (\bibinfo {year} {2022})}\BibitemShut {NoStop}%
\bibitem [{\citenamefont {Milchberg}\ \emph {et~al.}(2001)\citenamefont
  {Milchberg}, \citenamefont {McNaught},\ and\ \citenamefont
  {Parra}}]{Milchberg2001PRE}%
  \BibitemOpen
  \bibfield  {author} {\bibinfo {author} {\bibfnamefont {H.~M.}\ \bibnamefont
  {Milchberg}}, \bibinfo {author} {\bibfnamefont {S.~J.}\ \bibnamefont
  {McNaught}},\ and\ \bibinfo {author} {\bibfnamefont {E.}~\bibnamefont
  {Parra}},\ }\bibfield  {title} {\bibinfo {title} {Plasma hydrodynamics of the
  intense laser-cluster interaction},\ }\href
  {https://doi.org/10.1103/physreve.64.056402} {\bibfield  {journal} {\bibinfo
  {journal} {Phys. Rev. E}\ }\textbf {\bibinfo {volume} {64}},\ \bibinfo
  {pages} {056402} (\bibinfo {year} {2001})}\BibitemShut {NoStop}%
\bibitem [{\citenamefont {K\:{o}ller}\ \emph {et~al.}(1999)\citenamefont
  {K\:{o}ller}, \citenamefont {Schumacher}, \citenamefont {K\:{o}hn},
  \citenamefont {Teuber}, \citenamefont {Tiggesb\:{a}umker},\ and\
  \citenamefont {Meiwes-Broer}}]{Koller1999PRL}%
  \BibitemOpen
  \bibfield  {author} {\bibinfo {author} {\bibfnamefont {L.}~\bibnamefont
  {K\:{o}ller}}, \bibinfo {author} {\bibfnamefont {M.}~\bibnamefont
  {Schumacher}}, \bibinfo {author} {\bibfnamefont {J.}~\bibnamefont
  {K\:{o}hn}}, \bibinfo {author} {\bibfnamefont {S.}~\bibnamefont {Teuber}},
  \bibinfo {author} {\bibfnamefont {J.}~\bibnamefont {Tiggesb\:{a}umker}},\
  and\ \bibinfo {author} {\bibfnamefont {K.~H.}\ \bibnamefont {Meiwes-Broer}},\
  }\bibfield  {title} {\bibinfo {title} {Plasmon-enhanced multi-ionization of
  small metal clusters in strong femtosecond laser fields},\ }\href
  {https://doi.org/10.1103/PhysRevLett.82.3783} {\bibfield  {journal} {\bibinfo
   {journal} {Phys. Rev. Lett.}\ }\textbf {\bibinfo {volume} {82}},\ \bibinfo
  {pages} {3783} (\bibinfo {year} {1999})}\BibitemShut {NoStop}%
\bibitem [{\citenamefont {Jungreuthmayer}\ \emph {et~al.}(2004)\citenamefont
  {Jungreuthmayer}, \citenamefont {Geissler}, \citenamefont {Zanghellini},\
  and\ \citenamefont {Brabec}}]{Jungreuthmayer2004PRL}%
  \BibitemOpen
  \bibfield  {author} {\bibinfo {author} {\bibfnamefont {C.}~\bibnamefont
  {Jungreuthmayer}}, \bibinfo {author} {\bibfnamefont {M.}~\bibnamefont
  {Geissler}}, \bibinfo {author} {\bibfnamefont {J.}~\bibnamefont
  {Zanghellini}},\ and\ \bibinfo {author} {\bibfnamefont {T.}~\bibnamefont
  {Brabec}},\ }\bibfield  {title} {\bibinfo {title} {Microscopic analysis of
  large-cluster explosion in intense laser fields},\ }\href
  {https://doi.org/10.1103/PhysRevLett.92.133401} {\bibfield  {journal}
  {\bibinfo  {journal} {Phys. Rev. Lett.}\ }\textbf {\bibinfo {volume} {92}},\
  \bibinfo {pages} {133401} (\bibinfo {year} {2004})}\BibitemShut {NoStop}%
\bibitem [{\citenamefont {Psikal}\ \emph {et~al.}(2011)\citenamefont {Psikal},
  \citenamefont {Klimo},\ and\ \citenamefont {Limpouch}}]{Psikal2011NIMPRA}%
  \BibitemOpen
  \bibfield  {author} {\bibinfo {author} {\bibfnamefont {J.}~\bibnamefont
  {Psikal}}, \bibinfo {author} {\bibfnamefont {O.}~\bibnamefont {Klimo}},\ and\
  \bibinfo {author} {\bibfnamefont {J.}~\bibnamefont {Limpouch}},\ }\bibfield
  {title} {\bibinfo {title} {Field ionization effects on ion acceleration in
  laser-irradiated clusters},\ }\href
  {https://doi.org/10.1016/j.nima.2011.01.068} {\bibfield  {journal} {\bibinfo
  {journal} {Nucl. Instrum. Methods. Phys. Res. A}\ }\textbf {\bibinfo {volume}
  {653}},\ \bibinfo {pages} {109} (\bibinfo {year} {2011})}\BibitemShut
  {NoStop}%
\bibitem [{\citenamefont {Bystrov}\ and\ \citenamefont
  {Gildenburg}(2009)}]{Bystrov2009PRL}%
  \BibitemOpen
  \bibfield  {author} {\bibinfo {author} {\bibfnamefont {A.~M.}\ \bibnamefont
  {Bystrov}}\ and\ \bibinfo {author} {\bibfnamefont {V.~B.}\ \bibnamefont
  {Gildenburg}},\ }\bibfield  {title} {\bibinfo {title} {Infrared to
  ultraviolet light conversion in laser-cluster interactions},\ }\href
  {https://doi.org/10.1103/PhysRevLett.103.083401} {\bibfield  {journal}
  {\bibinfo  {journal} {Phys. Rev. Lett.}\ }\textbf {\bibinfo {volume} {103}},\
  \bibinfo {pages} {083401} (\bibinfo {year} {2009})}\BibitemShut {NoStop}%
\bibitem [{\citenamefont {Gao}\ \emph {et~al.}(2019)\citenamefont {Gao},
  \citenamefont {Shim},\ and\ \citenamefont {Downer}}]{Gao2019OL}%
  \BibitemOpen
  \bibfield  {author} {\bibinfo {author} {\bibfnamefont {X.}~\bibnamefont
  {Gao}}, \bibinfo {author} {\bibfnamefont {B.}~\bibnamefont {Shim}},\ and\
  \bibinfo {author} {\bibfnamefont {M.~C.}\ \bibnamefont {Downer}},\ }\bibfield
   {title} {\bibinfo {title} {Brunel harmonics generated from ionizing clusters
  by few-cycle laser pulses},\ }\href
  {http://ol.osa.org/abstract.cfm?URI=ol-44-4-779} {\bibfield  {journal}
  {\bibinfo  {journal} {Opt. Lett.}\ }\textbf {\bibinfo {volume} {44}},\
  \bibinfo {pages} {779} (\bibinfo {year} {2019})}\BibitemShut {NoStop}%
\bibitem [{\citenamefont {McCormick}\ \emph {et~al.}(2014)\citenamefont
  {McCormick}, \citenamefont {Arefiev}, \citenamefont {Quevedo}, \citenamefont
  {Bengtson},\ and\ \citenamefont {Ditmire}}]{McCormick2014PRL}%
  \BibitemOpen
  \bibfield  {author} {\bibinfo {author} {\bibfnamefont {M.}~\bibnamefont
  {McCormick}}, \bibinfo {author} {\bibfnamefont {A.~V.}\ \bibnamefont
  {Arefiev}}, \bibinfo {author} {\bibfnamefont {H.~J.}\ \bibnamefont
  {Quevedo}}, \bibinfo {author} {\bibfnamefont {R.~D.}\ \bibnamefont
  {Bengtson}},\ and\ \bibinfo {author} {\bibfnamefont {T.}~\bibnamefont
  {Ditmire}},\ }\bibfield  {title} {\bibinfo {title} {Observation of
  self-sustaining relativistic ionization wave launched by a sheath field},\
  }\href {https://doi.org/10.1103/PhysRevLett.112.045002} {\bibfield  {journal}
  {\bibinfo  {journal} {Phys. Rev. Lett.}\ }\textbf {\bibinfo {volume} {112}},\
  \bibinfo {pages} {045002} (\bibinfo {year} {2014})}\BibitemShut {NoStop}%
\bibitem [{\citenamefont {Ditmire}\ \emph {et~al.}(1996)\citenamefont
  {Ditmire}, \citenamefont {Donnelly}, \citenamefont {Rubenchik}, \citenamefont
  {Falcone},\ and\ \citenamefont {Perry}}]{Ditmire1996PRA}%
  \BibitemOpen
  \bibfield  {author} {\bibinfo {author} {\bibfnamefont {T.}~\bibnamefont
  {Ditmire}}, \bibinfo {author} {\bibfnamefont {T.}~\bibnamefont {Donnelly}},
  \bibinfo {author} {\bibfnamefont {A.~M.}\ \bibnamefont {Rubenchik}}, \bibinfo
  {author} {\bibfnamefont {R.~W.}\ \bibnamefont {Falcone}},\ and\ \bibinfo
  {author} {\bibfnamefont {M.~D.}\ \bibnamefont {Perry}},\ }\bibfield  {title}
  {\bibinfo {title} {Interaction of intense laser pulses with atomic
  clusters},\ }\href@noop {} {\bibfield  {journal} {\bibinfo  {journal} {Phys.
  Rev. A}\ }\textbf {\bibinfo {volume} {53}},\ \bibinfo {pages} {3379}
  (\bibinfo {year} {1996})}\BibitemShut {NoStop}%
\bibitem [{\citenamefont {Fourment}\ \emph {et~al.}(2018)\citenamefont
  {Fourment}, \citenamefont {Chimier}, \citenamefont {Deneuville},
  \citenamefont {Descamps}, \citenamefont {Dorchies}, \citenamefont
  {Duchateau}, \citenamefont {Nadeau},\ and\ \citenamefont
  {Petit}}]{Fourment2018PRB}%
  \BibitemOpen
  \bibfield  {author} {\bibinfo {author} {\bibfnamefont {C.}~\bibnamefont
  {Fourment}}, \bibinfo {author} {\bibfnamefont {B.}~\bibnamefont {Chimier}},
  \bibinfo {author} {\bibfnamefont {F.}~\bibnamefont {Deneuville}}, \bibinfo
  {author} {\bibfnamefont {D.}~\bibnamefont {Descamps}}, \bibinfo {author}
  {\bibfnamefont {F.}~\bibnamefont {Dorchies}}, \bibinfo {author}
  {\bibfnamefont {G.}~\bibnamefont {Duchateau}}, \bibinfo {author}
  {\bibfnamefont {M.-C.}\ \bibnamefont {Nadeau}},\ and\ \bibinfo {author}
  {\bibfnamefont {S.}~\bibnamefont {Petit}},\ }\bibfield  {title} {\bibinfo
  {title} {Ultrafast changes in optical properties of ${\mathrm{sio}}_{2}$
  excited by femtosecond laser at the damage threshold and above},\ }\href
  {https://doi.org/10.1103/PhysRevB.98.155110} {\bibfield  {journal} {\bibinfo
  {journal} {Phys. Rev. B}\ }\textbf {\bibinfo {volume} {98}},\ \bibinfo
  {pages} {155110} (\bibinfo {year} {2018})}\BibitemShut {NoStop}%
\bibitem [{\citenamefont {Saalmann}\ \emph {et~al.}(2006)\citenamefont
  {Saalmann}, \citenamefont {Siedschlag},\ and\ \citenamefont
  {Rost}}]{Saalmann2006JPB}%
  \BibitemOpen
  \bibfield  {author} {\bibinfo {author} {\bibfnamefont {U.}~\bibnamefont
  {Saalmann}}, \bibinfo {author} {\bibfnamefont {C.}~\bibnamefont
  {Siedschlag}},\ and\ \bibinfo {author} {\bibfnamefont {J.~M.}\ \bibnamefont
  {Rost}},\ }\bibfield  {title} {\bibinfo {title} {Mechanisms of cluster
  ionization in strong laser pulses},\ }\href
  {http://stacks.iop.org/0953-4075/39/i=4/a=R01} {\bibfield  {journal}
  {\bibinfo  {journal} {J. Phys. B}\ }\textbf {\bibinfo {volume} {39}},\
  \bibinfo {pages} {R39} (\bibinfo {year} {2006})}\BibitemShut {NoStop}%
\bibitem [{\citenamefont {Park}\ \emph {et~al.}(2022)\citenamefont {Park},
  \citenamefont {Camacho~Garibay}, \citenamefont {Wang}, \citenamefont
  {Gorman}, \citenamefont {Agostini},\ and\ \citenamefont
  {DiMauro}}]{Park2022PRL}%
  \BibitemOpen
  \bibfield  {author} {\bibinfo {author} {\bibfnamefont {H.}~\bibnamefont
  {Park}}, \bibinfo {author} {\bibfnamefont {A.}~\bibnamefont
  {Camacho~Garibay}}, \bibinfo {author} {\bibfnamefont {Z.}~\bibnamefont
  {Wang}}, \bibinfo {author} {\bibfnamefont {T.}~\bibnamefont {Gorman}},
  \bibinfo {author} {\bibfnamefont {P.}~\bibnamefont {Agostini}},\ and\
  \bibinfo {author} {\bibfnamefont {L.~F.}\ \bibnamefont {DiMauro}},\
  }\bibfield  {title} {\bibinfo {title} {Unveiling the inhomogeneous nature of
  strong field ionization in extended systems},\ }\href
  {https://doi.org/10.1103/PhysRevLett.129.203202} {\bibfield  {journal}
  {\bibinfo  {journal} {Phys. Rev. Lett.}\ }\textbf {\bibinfo {volume} {129}},\
  \bibinfo {pages} {203202} (\bibinfo {year} {2022})}\BibitemShut {NoStop}%
\bibitem [{\citenamefont {Taguchi}\ \emph {et~al.}(2004)\citenamefont
  {Taguchi}, \citenamefont {Antonsen},\ and\ \citenamefont
  {Milchberg}}]{Taguchi2004PRL}%
  \BibitemOpen
  \bibfield  {author} {\bibinfo {author} {\bibfnamefont {T.}~\bibnamefont
  {Taguchi}}, \bibinfo {author} {\bibfnamefont {T.~M.}\ \bibnamefont
  {Antonsen}},\ and\ \bibinfo {author} {\bibfnamefont {H.~M.}\ \bibnamefont
  {Milchberg}},\ }\bibfield  {title} {\bibinfo {title} {Resonant heating of a
  cluster plasma by intense laser light},\ }\href
  {https://doi.org/10.1103/PhysRevLett.92.205003} {\bibfield  {journal}
  {\bibinfo  {journal} {Phys. Rev. Lett.}\ }\textbf {\bibinfo {volume} {92}},\
  \bibinfo {pages} {205003} (\bibinfo {year} {2004})}\BibitemShut {NoStop}%
\bibitem [{\citenamefont {Breizman}\ \emph {et~al.}(2005)\citenamefont
  {Breizman}, \citenamefont {Arefiev},\ and\ \citenamefont
  {Fomyts'kyi}}]{Breizman2005PP}%
  \BibitemOpen
  \bibfield  {author} {\bibinfo {author} {\bibfnamefont {B.~N.}\ \bibnamefont
  {Breizman}}, \bibinfo {author} {\bibfnamefont {A.~V.}\ \bibnamefont
  {Arefiev}},\ and\ \bibinfo {author} {\bibfnamefont {M.~V.}\ \bibnamefont
  {Fomyts'kyi}},\ }\bibfield  {title} {\bibinfo {title} {Nonlinear physics of
  laser-irradiated microclusters},\ }\href {https://doi.org/10.1063/1.1871939}
  {\bibfield  {journal} {\bibinfo  {journal} {Phys. Plasmas}\ }\textbf
  {\bibinfo {volume} {12}},\ \bibinfo {pages} {056706} (\bibinfo {year}
  {2005})}\BibitemShut {NoStop}%
\bibitem [{\citenamefont {Brunel}(1987)}]{Brunel1987PRL}%
  \BibitemOpen
  \bibfield  {author} {\bibinfo {author} {\bibfnamefont {F.}~\bibnamefont
  {Brunel}},\ }\bibfield  {title} {\bibinfo {title} {Not-so-resonant, resonant
  absorption},\ }\href {https://doi.org/10.1103/PhysRevLett.59.52} {\bibfield
  {journal} {\bibinfo  {journal} {Phys. Rev. Lett.}\ }\textbf {\bibinfo
  {volume} {59}},\ \bibinfo {pages} {52} (\bibinfo {year} {1987})}\BibitemShut
  {NoStop}%
\bibitem [{\citenamefont {Quere}\ \emph {et~al.}(2006)\citenamefont {Quere},
  \citenamefont {Thaury}, \citenamefont {Monot}, \citenamefont {Dobosz},
  \citenamefont {Martin}, \citenamefont {Geindre},\ and\ \citenamefont
  {Audebert}}]{Quere2006PRL}%
  \BibitemOpen
  \bibfield  {author} {\bibinfo {author} {\bibfnamefont {F.}~\bibnamefont
  {Quere}}, \bibinfo {author} {\bibfnamefont {C.}~\bibnamefont {Thaury}},
  \bibinfo {author} {\bibfnamefont {P.}~\bibnamefont {Monot}}, \bibinfo
  {author} {\bibfnamefont {S.}~\bibnamefont {Dobosz}}, \bibinfo {author}
  {\bibfnamefont {P.}~\bibnamefont {Martin}}, \bibinfo {author} {\bibfnamefont
  {J.~P.}\ \bibnamefont {Geindre}},\ and\ \bibinfo {author} {\bibfnamefont
  {P.}~\bibnamefont {Audebert}},\ }\bibfield  {title} {\bibinfo {title}
  {Coherent wake emission of high-order harmonics from overdense plasmas},\
  }\bibfield  {journal} {\bibinfo  {journal} {Phys. Rev. Lett.}\ }\textbf
  {\bibinfo {volume} {96}},\ \href
  {https://doi.org/10.1103/PhysRevLett.96.125004}
  {10.1103/PhysRevLett.96.125004} (\bibinfo {year} {2006})\BibitemShut
  {NoStop}%
\bibitem [{\citenamefont {Thaury}\ and\ \citenamefont
  {Qu\'{e}r\'{e}}(2010)}]{Thaury2010JPBAMOP}%
  \BibitemOpen
  \bibfield  {author} {\bibinfo {author} {\bibfnamefont {C.}~\bibnamefont
  {Thaury}}\ and\ \bibinfo {author} {\bibfnamefont {F.}~\bibnamefont
  {Qu\'{e}r\'{e}}},\ }\bibfield  {title} {\bibinfo {title} {High-order harmonic
  and attosecond pulse generation on plasma mirrors: basic mechanisms},\ }\href
  {https://doi.org/10.1088/0953-4075/43/21/213001} {\bibfield  {journal}
  {\bibinfo  {journal} {J. Phys. B: At., Mol. Opt. Phys.}\ }\textbf {\bibinfo
  {volume} {43}},\ \bibinfo {pages} {213001} (\bibinfo {year}
  {2010})}\BibitemShut {NoStop}%
\bibitem [{\citenamefont {Derouillat}\ \emph {et~al.}(2018)\citenamefont
  {Derouillat}, \citenamefont {Beck}, \citenamefont {P{\'e}rez}, \citenamefont
  {Vinci}, \citenamefont {Chiaramello}, \citenamefont {Grassi}, \citenamefont
  {Fl{\'e}}, \citenamefont {Bouchard}, \citenamefont {Plotnikov}, \citenamefont
  {Aunai}, \citenamefont {Dargent}, \citenamefont {Riconda},\ and\
  \citenamefont {Grech}}]{Derouillat2018CPC}%
  \BibitemOpen
  \bibfield  {author} {\bibinfo {author} {\bibfnamefont {J.}~\bibnamefont
  {Derouillat}}, \bibinfo {author} {\bibfnamefont {A.}~\bibnamefont {Beck}},
  \bibinfo {author} {\bibfnamefont {F.}~\bibnamefont {P{\'e}rez}}, \bibinfo
  {author} {\bibfnamefont {T.}~\bibnamefont {Vinci}}, \bibinfo {author}
  {\bibfnamefont {M.}~\bibnamefont {Chiaramello}}, \bibinfo {author}
  {\bibfnamefont {A.}~\bibnamefont {Grassi}}, \bibinfo {author} {\bibfnamefont
  {M.}~\bibnamefont {Fl{\'e}}}, \bibinfo {author} {\bibfnamefont
  {G.}~\bibnamefont {Bouchard}}, \bibinfo {author} {\bibfnamefont
  {I.}~\bibnamefont {Plotnikov}}, \bibinfo {author} {\bibfnamefont
  {N.}~\bibnamefont {Aunai}}, \bibinfo {author} {\bibfnamefont
  {J.}~\bibnamefont {Dargent}}, \bibinfo {author} {\bibfnamefont
  {C.}~\bibnamefont {Riconda}},\ and\ \bibinfo {author} {\bibfnamefont
  {M.}~\bibnamefont {Grech}},\ }\bibfield  {title} {\bibinfo {title} {Smilei :
  A collaborative, open-source, multi-purpose particle-in-cell code for plasma
  simulation},\ }\href {https://doi.org/10.1016/j.cpc.2017.09.024} {\bibfield
  {journal} {\bibinfo  {journal} {Comput. Phys. Commun.}\ }\textbf {\bibinfo
  {volume} {222}},\ \bibinfo {pages} {351} (\bibinfo {year}
  {2018})}\BibitemShut {NoStop}%
\bibitem [{\citenamefont {Gao}(2023)}]{Gao2023PP}%
  \BibitemOpen
  \bibfield  {author} {\bibinfo {author} {\bibfnamefont {X.}~\bibnamefont
  {Gao}},\ }\bibfield  {title} {\bibinfo {title} {{Ionization dynamics of
  sub-micrometer-sized clusters in intense ultrafast laser pulses}},\ }\href
  {https://doi.org/10.1063/5.0143356} {\bibfield  {journal} {\bibinfo
  {journal} {Phys. Plasmas}\ }\textbf {\bibinfo {volume} {30}},\ \bibinfo
  {pages} {052102} (\bibinfo {year} {2023})}\BibitemShut {NoStop}%
\bibitem [{\citenamefont {Gao}\ \emph {et~al.}(2013)\citenamefont {Gao},
  \citenamefont {Arefiev}, \citenamefont {Korzekwa}, \citenamefont {Wang},
  \citenamefont {Shim},\ and\ \citenamefont {Downer}}]{Gao2013JAP}%
  \BibitemOpen
  \bibfield  {author} {\bibinfo {author} {\bibfnamefont {X.}~\bibnamefont
  {Gao}}, \bibinfo {author} {\bibfnamefont {A.~V.}\ \bibnamefont {Arefiev}},
  \bibinfo {author} {\bibfnamefont {R.~C.}\ \bibnamefont {Korzekwa}}, \bibinfo
  {author} {\bibfnamefont {X.}~\bibnamefont {Wang}}, \bibinfo {author}
  {\bibfnamefont {B.}~\bibnamefont {Shim}},\ and\ \bibinfo {author}
  {\bibfnamefont {M.~C.}\ \bibnamefont {Downer}},\ }\bibfield  {title}
  {\bibinfo {title} {Spatio-temporal profiling of cluster mass fraction in a
  pulsed supersonic gas jet by frequency-domain holography},\ }\href
  {https://doi.org/http://dx.doi.org/10.1063/1.4815961} {\bibfield  {journal}
  {\bibinfo  {journal} {J. Appl. Phys.}\ }\textbf {\bibinfo {volume} {114}},\
  \bibinfo {pages} {034903} (\bibinfo {year} {2013})}\BibitemShut {NoStop}%
\bibitem [{\citenamefont {Corkum}(1993)}]{Corkum1993PRL}%
  \BibitemOpen
  \bibfield  {author} {\bibinfo {author} {\bibfnamefont {P.~B.}\ \bibnamefont
  {Corkum}},\ }\bibfield  {title} {\bibinfo {title} {Plasma perspective on
  strong-field multiphoton ionization},\ }\href
  {https://doi.org/10.1103/PhysRevLett.71.1994} {\bibfield  {journal} {\bibinfo
   {journal} {Phys. Rev. Lett.}\ }\textbf {\bibinfo {volume} {71}},\ \bibinfo
  {pages} {1994} (\bibinfo {year} {1993})}\BibitemShut {NoStop}%
\bibitem [{\citenamefont {Liseykina}\ and\ \citenamefont
  {Bauer}(2013)}]{Liseykina2013PRL}%
  \BibitemOpen
  \bibfield  {author} {\bibinfo {author} {\bibfnamefont {T.~V.}\ \bibnamefont
  {Liseykina}}\ and\ \bibinfo {author} {\bibfnamefont {D.}~\bibnamefont
  {Bauer}},\ }\bibfield  {title} {\bibinfo {title} {Plasma-formation dynamics
  in intense laser-droplet interaction},\ }\href
  {https://link.aps.org/doi/10.1103/PhysRevLett.110.145003} {\bibfield
  {journal} {\bibinfo  {journal} {Phys. Rev. Lett.}\ }\textbf {\bibinfo
  {volume} {110}},\ \bibinfo {pages} {145003} (\bibinfo {year}
  {2013})}\BibitemShut {NoStop}%
\bibitem [{\citenamefont {Zastrau}\ \emph {et~al.}(2015)\citenamefont
  {Zastrau}, \citenamefont {Sperling}, \citenamefont {Fortmann-Grote},
  \citenamefont {Becker}, \citenamefont {Bornath}, \citenamefont {Bredow},
  \citenamefont {D\"{o}ppner}, \citenamefont {Fennel}, \citenamefont
  {Fletcher}, \citenamefont {F\"{o}rster}, \citenamefont {G\"{o}de},
  \citenamefont {Gregori}, \citenamefont {Harmand}, \citenamefont {Hilbert},
  \citenamefont {Laarmann}, \citenamefont {Lee}, \citenamefont {Ma},
  \citenamefont {Meiwes-Broer}, \citenamefont {Mithen}, \citenamefont {Murphy},
  \citenamefont {Nakatsutsumi}, \citenamefont {Neumayer}, \citenamefont
  {Przystawik}, \citenamefont {Skruszewicz}, \citenamefont {Tiggesb\"{a}umker},
  \citenamefont {Toleikis}, \citenamefont {White}, \citenamefont {Glenzer},
  \citenamefont {Redmer},\ and\ \citenamefont
  {Tschentscher}}]{Zastrau2015JPBAMOP}%
  \BibitemOpen
  \bibfield  {author} {\bibinfo {author} {\bibfnamefont {U.}~\bibnamefont
  {Zastrau}}, \bibinfo {author} {\bibfnamefont {P.}~\bibnamefont {Sperling}},
  \bibinfo {author} {\bibfnamefont {C.}~\bibnamefont {Fortmann-Grote}},
  \bibinfo {author} {\bibfnamefont {A.}~\bibnamefont {Becker}}, \bibinfo
  {author} {\bibfnamefont {T.}~\bibnamefont {Bornath}}, \bibinfo {author}
  {\bibfnamefont {R.}~\bibnamefont {Bredow}}, \bibinfo {author} {\bibfnamefont
  {T.}~\bibnamefont {D\"{o}ppner}}, \bibinfo {author} {\bibfnamefont
  {T.}~\bibnamefont {Fennel}}, \bibinfo {author} {\bibfnamefont {L.~B.}\
  \bibnamefont {Fletcher}}, \bibinfo {author} {\bibfnamefont {E.}~\bibnamefont
  {F\"{o}rster}}, \bibinfo {author} {\bibfnamefont {S.}~\bibnamefont
  {G\"{o}de}}, \bibinfo {author} {\bibfnamefont {G.}~\bibnamefont {Gregori}},
  \bibinfo {author} {\bibfnamefont {M.}~\bibnamefont {Harmand}}, \bibinfo
  {author} {\bibfnamefont {V.}~\bibnamefont {Hilbert}}, \bibinfo {author}
  {\bibfnamefont {T.}~\bibnamefont {Laarmann}}, \bibinfo {author}
  {\bibfnamefont {H.~J.}\ \bibnamefont {Lee}}, \bibinfo {author} {\bibfnamefont
  {T.}~\bibnamefont {Ma}}, \bibinfo {author} {\bibfnamefont {K.~H.}\
  \bibnamefont {Meiwes-Broer}}, \bibinfo {author} {\bibfnamefont {J.~P.}\
  \bibnamefont {Mithen}}, \bibinfo {author} {\bibfnamefont {C.~D.}\
  \bibnamefont {Murphy}}, \bibinfo {author} {\bibfnamefont {M.}~\bibnamefont
  {Nakatsutsumi}}, \bibinfo {author} {\bibfnamefont {P.}~\bibnamefont
  {Neumayer}}, \bibinfo {author} {\bibfnamefont {A.}~\bibnamefont
  {Przystawik}}, \bibinfo {author} {\bibfnamefont {S.}~\bibnamefont
  {Skruszewicz}}, \bibinfo {author} {\bibfnamefont {J.}~\bibnamefont
  {Tiggesb\"{a}umker}}, \bibinfo {author} {\bibfnamefont {S.}~\bibnamefont
  {Toleikis}}, \bibinfo {author} {\bibfnamefont {T.~G.}\ \bibnamefont {White}},
  \bibinfo {author} {\bibfnamefont {S.~H.}\ \bibnamefont {Glenzer}}, \bibinfo
  {author} {\bibfnamefont {R.}~\bibnamefont {Redmer}},\ and\ \bibinfo {author}
  {\bibfnamefont {T.}~\bibnamefont {Tschentscher}},\ }\bibfield  {title}
  {\bibinfo {title} {Ultrafast electron kinetics in short pulse laser-driven
  dense hydrogen},\ }\href {http://dx.doi.org/10.1088/0953-4075/48/22/224004}
  {\bibfield  {journal} {\bibinfo  {journal} {J. Phys. B: At., Mol. Opt.
  Phys.}\ }\textbf {\bibinfo {volume} {48}},\ \bibinfo {pages} {224004}
  (\bibinfo {year} {2015})}\BibitemShut {NoStop}%
\bibitem [{\citenamefont {Varin}\ \emph {et~al.}(2012)\citenamefont {Varin},
  \citenamefont {Peltz}, \citenamefont {Brabec},\ and\ \citenamefont
  {Fennel}}]{Varin2012PRL}%
  \BibitemOpen
  \bibfield  {author} {\bibinfo {author} {\bibfnamefont {C.}~\bibnamefont
  {Varin}}, \bibinfo {author} {\bibfnamefont {C.}~\bibnamefont {Peltz}},
  \bibinfo {author} {\bibfnamefont {T.}~\bibnamefont {Brabec}},\ and\ \bibinfo
  {author} {\bibfnamefont {T.}~\bibnamefont {Fennel}},\ }\bibfield  {title}
  {\bibinfo {title} {Attosecond plasma wave dynamics in laser-driven cluster
  nanoplasmas},\ }\href {https://doi.org/10.1103/PhysRevLett.108.175007}
  {\bibfield  {journal} {\bibinfo  {journal} {Phys. Rev. Lett.}\ }\textbf
  {\bibinfo {volume} {108}},\ \bibinfo {pages} {175007} (\bibinfo {year}
  {2012})}\BibitemShut {NoStop}%
\bibitem [{\citenamefont {Mikaberidze}\ \emph {et~al.}(2009)\citenamefont
  {Mikaberidze}, \citenamefont {Saalmann},\ and\ \citenamefont
  {Rost}}]{Mikaberidze2009PRL}%
  \BibitemOpen
  \bibfield  {author} {\bibinfo {author} {\bibfnamefont {A.}~\bibnamefont
  {Mikaberidze}}, \bibinfo {author} {\bibfnamefont {U.}~\bibnamefont
  {Saalmann}},\ and\ \bibinfo {author} {\bibfnamefont {J.~M.}\ \bibnamefont
  {Rost}},\ }\bibfield  {title} {\bibinfo {title} {Laser-driven nanoplasmas in
  doped helium droplets: Local ignition and anisotropic growth},\ }\href
  {https://link.aps.org/doi/10.1103/PhysRevLett.102.128102} {\bibfield
  {journal} {\bibinfo  {journal} {Phys. Rev. Lett.}\ }\textbf {\bibinfo
  {volume} {102}},\ \bibinfo {pages} {128102} (\bibinfo {year}
  {2009})}\BibitemShut {NoStop}%
\bibitem [{\citenamefont {Gorkhover}\ \emph {et~al.}(2016)\citenamefont
  {Gorkhover}, \citenamefont {Schorb}, \citenamefont {Coffee}, \citenamefont
  {Adolph}, \citenamefont {Foucar}, \citenamefont {Rupp}, \citenamefont
  {Aquila}, \citenamefont {Bozek}, \citenamefont {Epp}, \citenamefont {Erk},
  \citenamefont {Gumprecht}, \citenamefont {Holmegaard}, \citenamefont
  {Hartmann}, \citenamefont {Hartmann}, \citenamefont {Hauser}, \citenamefont
  {Holl}, \citenamefont {H\"{o}mke}, \citenamefont {Johnsson}, \citenamefont
  {Kimmel}, \citenamefont {K\"{u}hnel}, \citenamefont {Messerschmidt},
  \citenamefont {Reich}, \citenamefont {Rouz\'{e}e}, \citenamefont {Rudek},
  \citenamefont {Schmidt}, \citenamefont {Schulz}, \citenamefont {Soltau},
  \citenamefont {Stern}, \citenamefont {Weidenspointner}, \citenamefont
  {White}, \citenamefont {K\"{u}pper}, \citenamefont {Str\"{u}der},
  \citenamefont {Schlichting}, \citenamefont {Ullrich}, \citenamefont {Rolles},
  \citenamefont {Rudenko}, \citenamefont {M\"{o}ller},\ and\ \citenamefont
  {Bostedt}}]{Gorkhover2016NP}%
  \BibitemOpen
  \bibfield  {author} {\bibinfo {author} {\bibfnamefont {T.}~\bibnamefont
  {Gorkhover}}, \bibinfo {author} {\bibfnamefont {S.}~\bibnamefont {Schorb}},
  \bibinfo {author} {\bibfnamefont {R.}~\bibnamefont {Coffee}}, \bibinfo
  {author} {\bibfnamefont {M.}~\bibnamefont {Adolph}}, \bibinfo {author}
  {\bibfnamefont {L.}~\bibnamefont {Foucar}}, \bibinfo {author} {\bibfnamefont
  {D.}~\bibnamefont {Rupp}}, \bibinfo {author} {\bibfnamefont {A.}~\bibnamefont
  {Aquila}}, \bibinfo {author} {\bibfnamefont {J.~D.}\ \bibnamefont {Bozek}},
  \bibinfo {author} {\bibfnamefont {S.~W.}\ \bibnamefont {Epp}}, \bibinfo
  {author} {\bibfnamefont {B.}~\bibnamefont {Erk}}, \bibinfo {author}
  {\bibfnamefont {L.}~\bibnamefont {Gumprecht}}, \bibinfo {author}
  {\bibfnamefont {L.}~\bibnamefont {Holmegaard}}, \bibinfo {author}
  {\bibfnamefont {A.}~\bibnamefont {Hartmann}}, \bibinfo {author}
  {\bibfnamefont {R.}~\bibnamefont {Hartmann}}, \bibinfo {author}
  {\bibfnamefont {G.}~\bibnamefont {Hauser}}, \bibinfo {author} {\bibfnamefont
  {P.}~\bibnamefont {Holl}}, \bibinfo {author} {\bibfnamefont {A.}~\bibnamefont
  {H\"{o}mke}}, \bibinfo {author} {\bibfnamefont {P.}~\bibnamefont {Johnsson}},
  \bibinfo {author} {\bibfnamefont {N.}~\bibnamefont {Kimmel}}, \bibinfo
  {author} {\bibfnamefont {K.-U.}\ \bibnamefont {K\"{u}hnel}}, \bibinfo
  {author} {\bibfnamefont {M.}~\bibnamefont {Messerschmidt}}, \bibinfo {author}
  {\bibfnamefont {C.}~\bibnamefont {Reich}}, \bibinfo {author} {\bibfnamefont
  {A.}~\bibnamefont {Rouz\'{e}e}}, \bibinfo {author} {\bibfnamefont
  {B.}~\bibnamefont {Rudek}}, \bibinfo {author} {\bibfnamefont
  {C.}~\bibnamefont {Schmidt}}, \bibinfo {author} {\bibfnamefont
  {J.}~\bibnamefont {Schulz}}, \bibinfo {author} {\bibfnamefont
  {H.}~\bibnamefont {Soltau}}, \bibinfo {author} {\bibfnamefont
  {S.}~\bibnamefont {Stern}}, \bibinfo {author} {\bibfnamefont
  {G.}~\bibnamefont {Weidenspointner}}, \bibinfo {author} {\bibfnamefont
  {B.}~\bibnamefont {White}}, \bibinfo {author} {\bibfnamefont
  {J.}~\bibnamefont {K\"{u}pper}}, \bibinfo {author} {\bibfnamefont
  {L.}~\bibnamefont {Str\"{u}der}}, \bibinfo {author} {\bibfnamefont
  {I.}~\bibnamefont {Schlichting}}, \bibinfo {author} {\bibfnamefont
  {J.}~\bibnamefont {Ullrich}}, \bibinfo {author} {\bibfnamefont
  {D.}~\bibnamefont {Rolles}}, \bibinfo {author} {\bibfnamefont
  {A.}~\bibnamefont {Rudenko}}, \bibinfo {author} {\bibfnamefont
  {T.}~\bibnamefont {M\"{o}ller}},\ and\ \bibinfo {author} {\bibfnamefont
  {C.}~\bibnamefont {Bostedt}},\ }\bibfield  {title} {\bibinfo {title}
  {Femtosecond and nanometre visualization of structural dynamics in
  superheated nanoparticles},\ }\href
  {http://dx.doi.org/10.1038/nphoton.2015.264} {\bibfield  {journal} {\bibinfo
  {journal} {Nat. Photonics}\ }\textbf {\bibinfo {volume} {10}},\ \bibinfo
  {pages} {93} (\bibinfo {year} {2016})}\BibitemShut {NoStop}%
\bibitem [{\citenamefont {Bacellar}\ \emph {et~al.}(2022)\citenamefont
  {Bacellar}, \citenamefont {Chatterley}, \citenamefont {Lackner},
  \citenamefont {Pemmaraju}, \citenamefont {Tanyag}, \citenamefont {Verma},
  \citenamefont {Bernando}, \citenamefont {O'Connell}, \citenamefont {Bucher},
  \citenamefont {Ferguson}, \citenamefont {Gorkhover}, \citenamefont {Coffee},
  \citenamefont {Coslovich}, \citenamefont {Ray}, \citenamefont {Osipov},
  \citenamefont {Neumark}, \citenamefont {Bostedt}, \citenamefont {Vilesov},\
  and\ \citenamefont {Gessner}}]{Bacellar2022PRL}%
  \BibitemOpen
  \bibfield  {author} {\bibinfo {author} {\bibfnamefont {C.}~\bibnamefont
  {Bacellar}}, \bibinfo {author} {\bibfnamefont {A.~S.}\ \bibnamefont
  {Chatterley}}, \bibinfo {author} {\bibfnamefont {F.}~\bibnamefont {Lackner}},
  \bibinfo {author} {\bibfnamefont {C.~D.}\ \bibnamefont {Pemmaraju}}, \bibinfo
  {author} {\bibfnamefont {R.~M.~P.}\ \bibnamefont {Tanyag}}, \bibinfo {author}
  {\bibfnamefont {D.}~\bibnamefont {Verma}}, \bibinfo {author} {\bibfnamefont
  {C.}~\bibnamefont {Bernando}}, \bibinfo {author} {\bibfnamefont {S.~M.~O.}\
  \bibnamefont {O'Connell}}, \bibinfo {author} {\bibfnamefont {M.}~\bibnamefont
  {Bucher}}, \bibinfo {author} {\bibfnamefont {K.~R.}\ \bibnamefont
  {Ferguson}}, \bibinfo {author} {\bibfnamefont {T.}~\bibnamefont {Gorkhover}},
  \bibinfo {author} {\bibfnamefont {R.~N.}\ \bibnamefont {Coffee}}, \bibinfo
  {author} {\bibfnamefont {G.}~\bibnamefont {Coslovich}}, \bibinfo {author}
  {\bibfnamefont {D.}~\bibnamefont {Ray}}, \bibinfo {author} {\bibfnamefont
  {T.}~\bibnamefont {Osipov}}, \bibinfo {author} {\bibfnamefont {D.~M.}\
  \bibnamefont {Neumark}}, \bibinfo {author} {\bibfnamefont {C.}~\bibnamefont
  {Bostedt}}, \bibinfo {author} {\bibfnamefont {A.~F.}\ \bibnamefont
  {Vilesov}},\ and\ \bibinfo {author} {\bibfnamefont {O.}~\bibnamefont
  {Gessner}},\ }\bibfield  {title} {\bibinfo {title} {Anisotropic surface
  broadening and core depletion during the evolution of a strong-field induced
  nanoplasma},\ }\href {https://doi.org/10.1103/PhysRevLett.129.073201}
  {\bibfield  {journal} {\bibinfo  {journal} {Phys. Rev. Lett.}\ }\textbf
  {\bibinfo {volume} {129}},\ \bibinfo {pages} {073201} (\bibinfo {year}
  {2022})}\BibitemShut {NoStop}%
\end{thebibliography}
\end{document}